\documentclass[aps,prl,twocolumn,showpacs,superscriptaddress]{revtex4-1}

\usepackage{amsfonts}
\usepackage{amsmath}
\usepackage{amssymb}
\usepackage{graphicx}
\usepackage{bm}

\begin{document}


\title{$c$-axis charge gap and its critical point in the heavily doped Ba(Fe$_{1-x}$Co$_x$)$_2$As$_2$}


\author{M.~A.~Tanatar}
\email[Corresponding author: ]{tanatar@ameslab.gov}
\affiliation{Ames Laboratory, Ames, Iowa 50011, USA}

\author{N.~Ni}

\affiliation{Ames Laboratory, Ames, Iowa 50011, USA}
\affiliation{Department of Physics and Astronomy, Iowa State University, Ames, Iowa 50011, USA }

\author{A.~Thaler}

\affiliation{Ames Laboratory, Ames, Iowa 50011, USA}
\affiliation{Department of Physics and Astronomy, Iowa State University, Ames, Iowa 50011, USA }

\author{S.~L.~Bud'ko}
\affiliation{Ames Laboratory, Ames, Iowa 50011, USA}
\affiliation{Department of Physics and Astronomy, Iowa State University, Ames, Iowa 50011, USA }

\author{P.~C.~Canfield}
\affiliation{Ames Laboratory, Ames, Iowa 50011, USA}
\affiliation{Department of Physics and Astronomy, Iowa State University, Ames, Iowa 50011, USA }

\author{R.~Prozorov}
\affiliation{Ames Laboratory, Ames, Iowa 50011, USA}
\affiliation{Department of Physics and Astronomy, Iowa State University, Ames, Iowa 50011, USA }

\date{\today}


\begin{abstract}

Temperature-dependent inter-plane resistivity, $\rho _c(T)$,  was used to characterise the normal state of the iron-arsenide superconductor Ba(Fe$_{1-x}$Co$_x$)$_2$As$_2$ over a broad doping range $0\leq x<0.50$. The data were compared with in-plane resistivity, $\rho _a(T)$, and magnetic susceptibility, $\chi  (T)$, taken in $H \bot c$, as well as Co NMR Knight shift, $^{59}K$, and spin relaxation rate, $1/T_1T$. The inter-plane resistivity data show a clear correlation with the NMR Knight shift, assigned to the formation of the pseudo-gap. Evolution of $\rho _c(T)$ with doping reveals two characteristic energy scales. The temperature of the cross-over from non-metallic, increasing on cooling, behavior of $\rho _c(T)$ at high-temperatures  to metallic behavior at low temperatures, $T^*$, correlates well with an anomaly in all three magnetic measurements. This characteristic temperature, equal to approximately 200~K in the parent compound, $x$=0, decreases with doping and vanishes near $x^* \approx$0.25. For doping levels $x \geq 0.166$, an additional feature appears above $T^*$, with metallic behavior of $\rho _c (T)$ found above the low-temperature resistivity increase. The characteristic temperature of this charge-gap formation, $T_{\rm CG}$, vanishes at $x_{\rm CG} \simeq$0.30, paving the way to metallic, $T$-linear, $\rho_c (T)$ close to $x_{\rm CG}$ and super-linear $T$-dependence for $x>x_{\rm CG}$. None of these features are evident in the in-plane resistivity $\rho_a(T)$. For doping levels $x <x_{\rm CG}$, $\chi (T)$ shows a known, anomalous, $T$-linear dependence, which disappears for $x>x_{\rm CG}$. These features are consistent with the existence of a uniaxial charge gap, accompanying formation of the magnetic pseudogap, and its critical suppression with doping.
The inferred $c$-axis charge gap reflects the three-dimensional character of the electronic structure and of the magnetism in the iron arsenides.
\end{abstract}

\pacs{74.70.Dd,72.15.-v,74.25.Jb}




\maketitle



The metallic state of the, until recently, only known  high temperature superconductors, the compounds based of Cu-O elements and frequently referred to as the cuprates \cite{BandM}, is characterized by a plethora of anomalies. At low doping levels, anomalous behaviours are found in the temperature-dependent resistivity, magnetization, NMR Knight shift and relaxation rate, as well as in spectroscopic data \cite{TS}. These behaviours are consistent with a decrease in the density of states at low temperatures, usually assigned with pseudogap formation. The phenomenology and $k$-space distribution of the pseudogap in the cuprates is now well established \cite{Kaminski}, however, its microscopic origin is still debated \cite{pseudogap_origin}. Main theories and experiments link it to two neighboring phases, an antiferromagnetic Mott-insulator,  with pseudogap arising due to exotic magnetism \cite{magnetism}, or to a superconducting phase, as an effect of the preformed superconducting pairs \cite{pairing}. The pseudogap is universally observed in both hole and electron \cite{greene} doped cuprates, though it is much more pronounced in the former.

Discovery of superconductivity with high critical temperatures in FeAs-based materials \cite{hosono}, breaking the monopoly of the cuprates, naturally raises the question about the common features of the two families \cite{Mazin-Nature}. It fuels the hopes that one day the enigmatic mechanism of high temperature superconductivity will be understood.
One of the important features to understand from such comparison, is a possible link between superconductivity and the pseudogap.

Features consistent with pseudogap are indeed observed in the hole doped RFeAsO \cite{PG1111_Hall_resistivity,PG1111_NMR,PG1111_Photoemission1,PG1111_Photoemission2,Sm_Russian} (R= rare earth, 1111 compounds in the following). A clearly decreased density of states is found in ARPES measurements in (Ba,K)Fe$_2$As$_2$ (BaK122 compounds in the following) \cite{ARPES_PGBaK}.  Because the parent compounds of iron pnictides are metals, the pseudogap here is believed to arise from nesting instability \cite{ishida}.

On the contrary, the experimental situation in electron doped Ba(Fe$_{1-x}$Co$_x$)$_2$As$_2$ (BaCo122 in the following) is less clear. NMR studies suggest the existence of a pseudogap over the whole doping range, from magnetically ordered parent compound to overdoped superconductor, with a characteristic temperature of  560~K $\pm$150~K at optimal doping \cite{BaCoNMRPG1}; ARPES found a tiny feature just above the superconducting $T_c$ \cite{ARPESBaCo}, whereas the in-plane resistivity does not reveal any pseudogap-like features \cite{NiNiCo} and is well described in a broad composition range by a sum of $T$-linear and $T^2$-contrinutions \cite{NDL}.

We have recently undertaken extensive anisotropic electrical resistivity measurements on parent and optimally doped Ba(Fe$_{1-x}$Co$_x$)$_2$As$_2$ \cite{anisotropy,anisotropy2,detwinning}. In addition to a small $ac$-anisotropy, we found different temperature dependencies of the in-plane and inter-plane electrical resistivity. Here we report a systematic study of the evolution of the inter-plane resistivity with doping. We show that the anomalies in the inter-plane resistivity reflect the existence of the enigmatic pseudogap state in BaCo122. Clear correlation with NMR measurements in BaCo122 as a function of doping, \cite{BaCoNMRPG2} and the lack of any associated features in the in-plane transport, suggest uniaxial symmetry of the pseudogap.

Tracking the evolution of the characteristic features of the temperature-dependent inter-plane resistivity with doping we found a critical concentration, $x_{\rm CG} \approx$0.30, beyond which the pseudogap features disapper. Our magnetization measurements show that this corresponds to the concentration at which the magnetic susceptibility loses its anomalous $T$-linear increase at high temperatures.  At the critical concentration, the $\rho _c (T)$ is very close to linear. This evolution of the inter-plane electrical resistivity suggests a (quantum) critical point \cite{Mathur,NDL} on the edge of the pseudogap state.

\section{Experimental}

Single crystals of BaFe$_2$As$_2$ doped with Co were grown from a starting load of metallic Ba, FeAs and CoAs, as described in detail elsewhere \cite{NiNiCo}. Crystals were thick platelets with sizes as big as 12$\times$8$\times$1 mm$^3$ and large faces corresponding to the tetragonal (001) plane.  The actual content of Co in the crystals was determined with wavelength dispersive electron probe microanalysis and is the $x$-value used throughout this text.

In our study of resistivity anisotropy in Ba(Fe$_{1-x}$Co$_x$)$_2$As$_2$, undoped $x$=0 \cite{anisotropy2} and optimally doped $x$=0.074 \cite{anisotropy}, we have found that special care must be taken for measurements in configurations with current along the tetragonal $c$-axis so as to avoid effects associated with the exfoliation of the samples. Cutting and shaping into transport samples inevitably introduces cracks, which affect the effective geometric factors of the samples and, in case the cracks are deep, can produce admixture of the in-plane resistivity component. A strong tendency to exfoliate prevents the cutting of samples with $c \gg a$. This limitation puts severe constraints on the measurement technique.

Samples for electrical resistivity measurements with current flow along the tetragonal $c$ axis ($\rho_c$) were cut into (0.3-0.7)$\times$(0.3-0.7)$\times$(0.1-0.5)mm$^3$ ($a\times b\times c$) slabs. All sample dimensions were measured with an accuracy of about 10\%. Contacts to the samples were made by attaching silver wires using ultrapure tin, resulting in an ultra low contact resistance (less than 10 $\mu \Omega$) \cite{SUST}. Measurements of $\rho_c$ were made in the two-probe sample configuration. Contacts were covering the whole $ab$ plane area of the $c$-axis samples. A four-probe scheme was used to measure the resistance down to the contact to the sample, i.e. the sum of the actual sample resistance $R_s$ and contact resistance $R_c$ was measured. Taking into account that $R_s \gg R_c$, contact resistance represents a minor correction of the order of 1 to 5\%. This can be directly seen for superconducting samples \cite{anisotropy,SUST,vortex} at temperatures $T<T_c$, where $R_s =$0 and the measured resistance represents $R_c$.

\begin{figure}
	
	\includegraphics[width=1.0\linewidth]{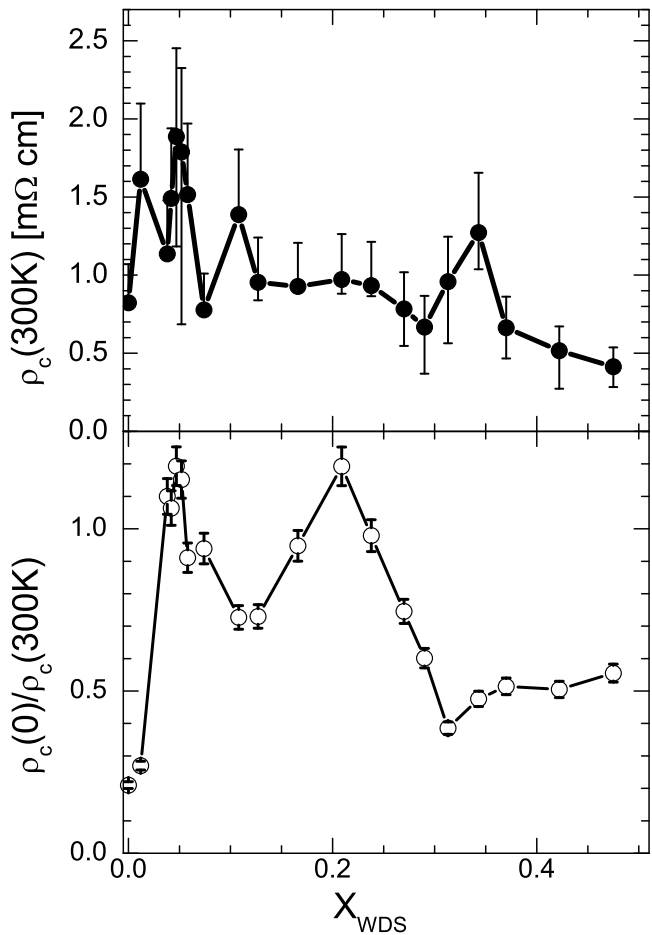}
	\caption{Room-temperature inter-plane resistivity, $\rho_c (300K)$, of Ba(Fe$_{1-x}$Co$_x$)$_2$As$_2$ as a function of doping (top panel). Lower  panel shows doping dependence of the ratio of resistivities at low temperatures and at room temperature, $\rho_c (0)/\rho_c (300K)$.  }
	\label{rhoc300K}
\end{figure}

The drawback of the measurement on samples with $c \gg a$ is that any inhomogeneity in the contact resistivity or internal sample connectivity admixes in-plane component due to redistribution of the current. To minimize this effect, we performed measurements of $\rho_c$ on at least 5 samples of each compositions. In all cases we obtained qualitatively similar temperature dependences of the electrical resistivity, as represented by the ratio of resistivities at room and low temperatures, $\rho _c (0)/\rho _c (300)$. The resistivity value, however, showed a notable scatter and at room temperature was typically in the range 1 to 2 m$\Omega$cm. For the sake of comparison we selected the samples with the temperature dependence of resistivity least similar to that of $\rho _a(T)$. The value of resistivity for these samples at room temperature is shown as a function of doping in the top panel of Fig.~\ref{rhoc300K}. Typically, these samples had the lowest value of electrical resistivity, as described in detail in Ref.~\onlinecite{anisotropy}. This is important since partial exfoliation increases resisitivity values \cite{anisotropy}. As a best demonstration of the correctness of our measurements, thermal conductivity measurements in the normal state for samples with $x$=0.127, accessed by the application of magnetic field, found Wiedemann-Franz law to be obeyed in $T \to $0 limit \cite{c-axis-thermalcond}. This {\it quantitative} coinsidence of two independent measurements is very important, because cracks can be partially transparent for phonons and thus affect thermal and electrical transport in a different way, leading to gross extrinsic Wiedemann-Franz law violation \cite{BETSGaCl4}. The evolutions of the inter-plane resistivity at room temperature, $\rho (300K)$, and of the residual resistivity ratio, $\rho_c (0)/ \rho(300K)$, with doping are summarized in Fig.~\ref{rhoc300K}. The resistivity value at room temperature for most compositions stays in the range 1 to 1.5 m$\Omega$cm, with doping it decreases to approximately 0.5 m$\Omega$cm. For several $x$ compositions we were not able to find crystals with resistivity values lower than 2~  m$\Omega$cm, despite the facts that (1) the evolution of the temperature-dependent resistivity for these samples followed the general trend, (2) close in $x$ compositions show usual resistivity values. This limits the accuracy of the absolute $\rho _c$ value determination  by approximately a factor of two.

\begin{figure}
	
	\includegraphics[width=1.0\linewidth]{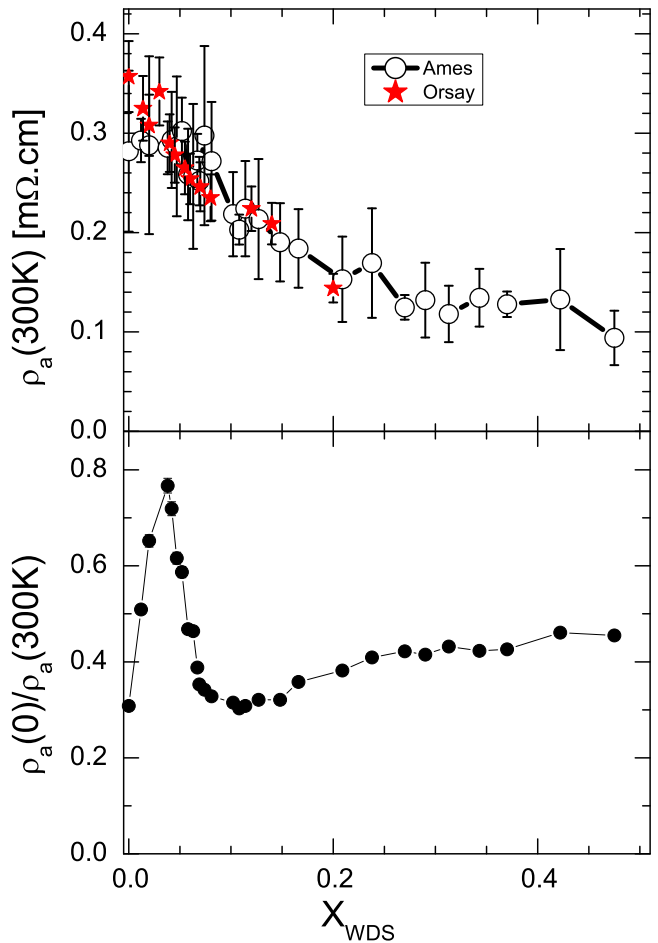}
	\caption{Room-temperature in-plane resistivity, $\rho_a (300K)$, of Ba(Fe$_{1-x}$Co$_x$)$_2$As$_2$ as a function of doping (top panel). Red stars show resistivity values taken from Ref.~\onlinecite{Alloul}. Lower  panel shows doping dependence of the resistivity ratio, $\rho_a (0)/\rho_a (300K)$.  }
	\label{rhoa300K}
\end{figure}

Samples for electrical resistivity measurements with current flow along the [100] $a$-axis in the tetragonal plane ($\rho _a$) were cut into bars of $(2-3) \times (0.1-0.2) \times (0.1-0.2)$ mm$^3$ ($a\times b\times c$).
Measurements of $\rho_a$ were made in both standard 4-probe and 2-probe configurations and gave identical results, see Ref.~[\onlinecite{SUST,vortex}]. Electrical resistivity of the samples at room temperature is shown as a function of doping in Fig.~\ref{rhoa300K}. Error bar represent statistisal standard deviation for at least 5 samples of each composition. The in-plane resistivity monotonically decreases from 270 $\mu \Omega$.cm in the parent compound to about 100 $\mu \Omega$.cm in the heavily overdoped composition with $x$=0.48. The magnitude of $\rho _a(300)$ is in good agreement with previous report over a narrower doping range \cite{Alloul}.  Residual resistivity ratio shows a rapid increase in the range where the Fermi surface topology change (Lifshits transition) happens (at $x \approx$0.025) \cite{mundeouk,Kaminskiivnature}, reaches a maximum at $x$=0.05 and then decreases towards minimum close to $x$=0.1. With further doping the ratio increases, the effect which mainly comes from a decrease of resistivity at room temperature.

The magnetization measurements were performed on cleaved samples to minimize the risk from small amount of surface flux. Samples typically had total mass of 10 to 20 mg. Measurements were performed in a standard MPSM SQUID magnetometer in a field of 5~T. Unless specially mentioned, magnetization measurements were performed in configuration $H \bot c$. For a composition $x$=0.325 measurements were also performed with $H \parallel c$. They found essentially no anisotropy, similar to our previous study \cite{NiNiCo}.

\section{Results}



\begin{figure}
		\includegraphics[width=0.9\linewidth]{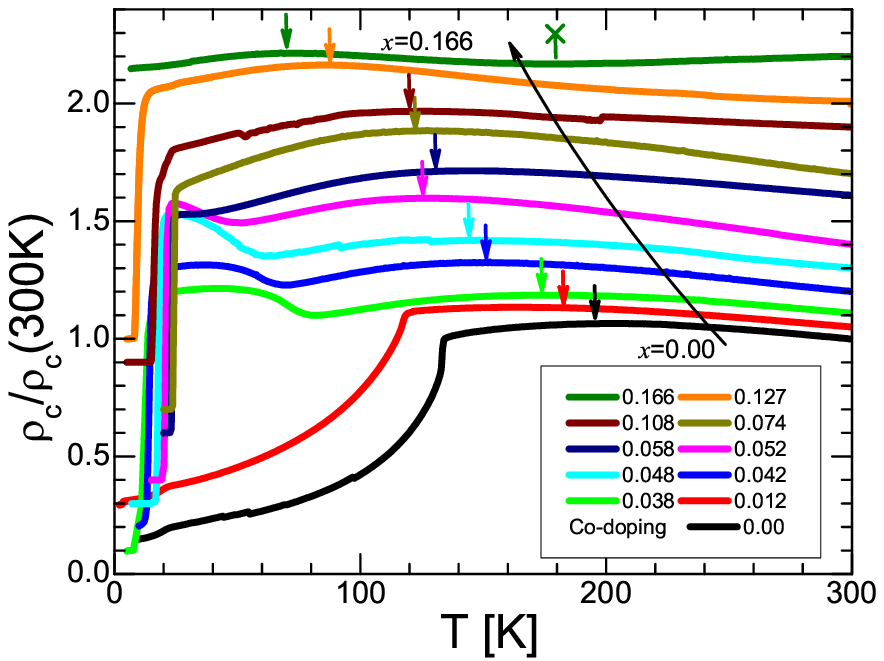}
		\includegraphics[width=0.9\linewidth]{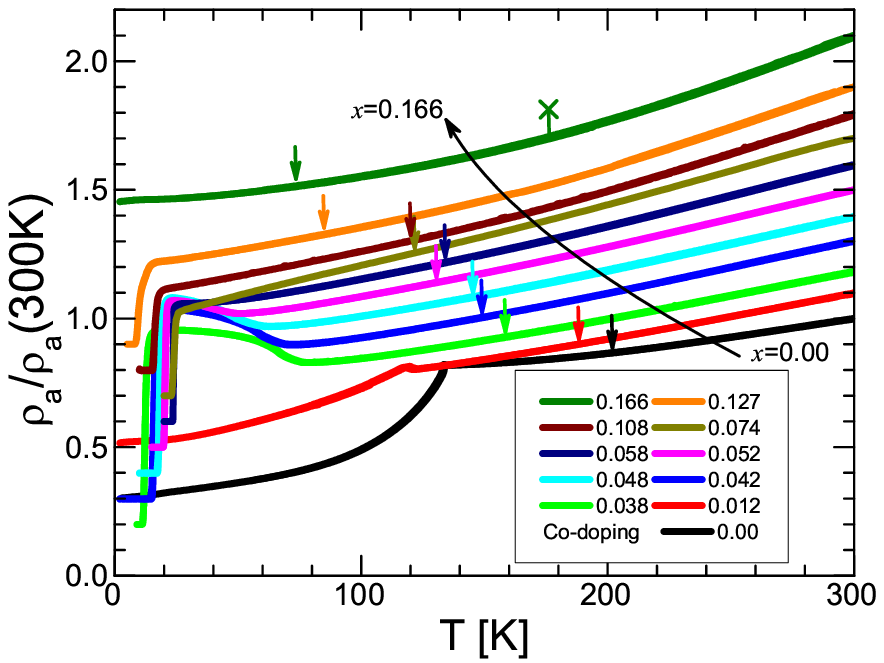}
	\caption{Temperature dependence of the inter-plane resistivity, $\rho_c $, normalized by its value at room temperature $\rho_c (300K)$, for samples of Ba(Fe$_{1-x}$Co$_x$)$_2$As$_2$ with $x \le 0.166$ (slightly above the concentration boundary for the superconducting dome) as shown in the figure (top panel). The curves are offset to avoid overlapping. Arrows show a position of the resistivity maximum, presented as a function of dopant concentration in the $T-x$ phase diagram (see Fig.~\ref{phaseDCo} below), cross-arrows show position of the resistivity minimum, $T_{\rm CG}$, appearing at high doping levels. Bottom panel shows doping evolution of the temperature-dependent in-plane resistivity, $\rho_a$, normalized by room-temperature value $\rho_a (300K)$. Arrows show positions of $T^*$ and $T_{\rm CG}$ as determined from $\rho _c(T)$, revealing no discernible features in the in-plane resistivity.
}
	\label{Co-rhoc-rhoa}
\end{figure}

\begin{figure}
		\includegraphics[width=0.9\linewidth]{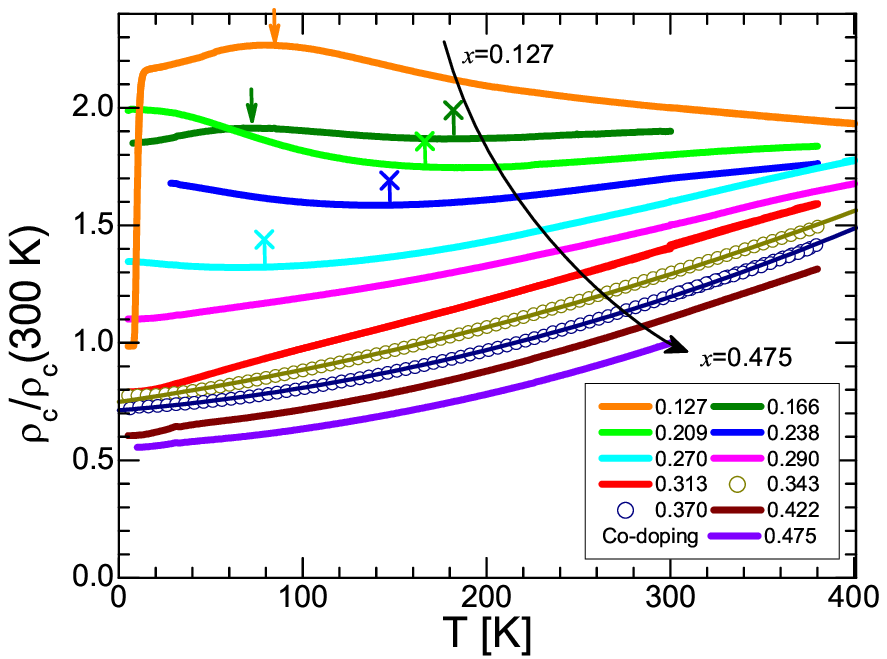}
		\includegraphics[width=0.9\linewidth]{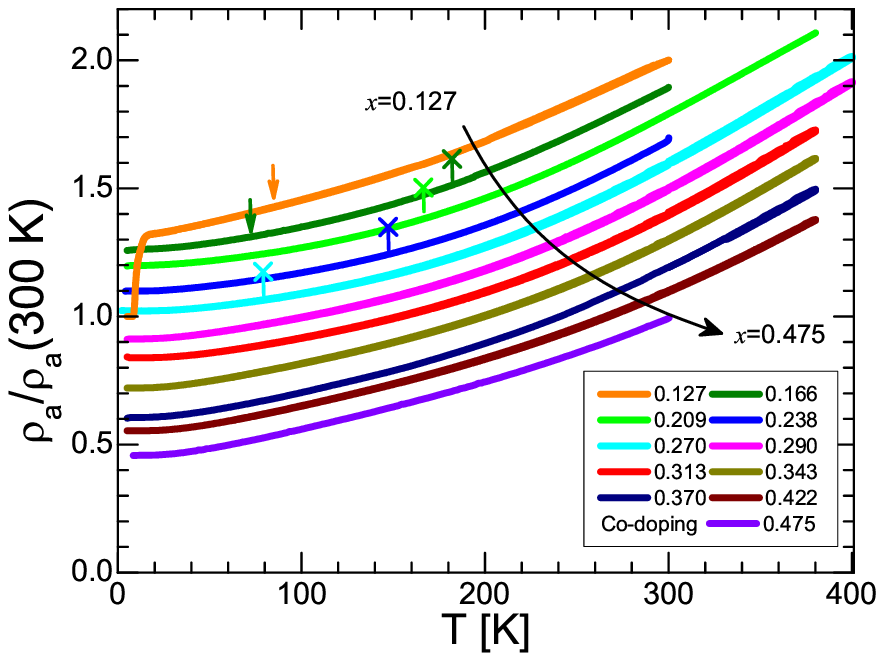}
	  \caption{Temperature dependence of the inter-plane resistivity, $\rho_c $, normalized by its value at room temperature $\rho_c (300K)$, for samples of Ba(Fe$_{1-x}$Co$_x$)$_2$As$_2$ with high doping levels $x \ge $0.127 as shown in the figure (top panel). The curves are offset to avoid overlapping. Cross-arrows and stright-arrows show positions of the resistivity minimum, $T_{\rm CG}$, and maximum, $T^*$, respectively. Bottom panel shows evolution of the temperature-dependent in-plane resistivity, $\rho_a$, normalized by room-temperature value $\rho_a (300K)$. Arrows show positions  of $T_{\rm CG}$ and  $T^*$ as determined from the inter-plane resistivity temperature dependence, revealing no discernible features in the in-plane resistivity.
}
	\label{Co-rhoc-rhoa_heavy}
\end{figure}

In Figs.~\ref{Co-rhoc-rhoa} and ~\ref{Co-rhoc-rhoa_heavy} we present evolution of the temperature-dependent resistivity with doping. The inter-plane resistivity (top panel, Fig.~\ref{Co-rhoc-rhoa}) of the parent compound decreases sharply below $T_{SM}$, similar to the in-plane resistivity (bottom panel, Fig.~\ref{Co-rhoc-rhoa}). In the inter-plane resistivity the decrease at $T_{SM}$=135~K is preceded with resistivity maximum at $T^* ~\approx$~200~K (shown with arrow in Fig.~\ref{Co-rhoc-rhoa}). With doping, the decrease of $\rho_c (T)$ below $T_{SM}$ turns into an increase (as seen for samples with $x$=0.038 to 0.058), similar to the behavior of $\rho _a (T)$, which shows two anomalies due to split structural/magnetic transition \cite{NiNiCo}. This change near $x \approx$0.025 is consistent with the proposed Lifshitz transition (Fermi surface topology change) as seen in thermoelectric power, Hall effect measurements \cite{mundeouk} and ARPES \cite{Kaminskiivnature}. However, the maximum in $\rho_c (T)$ at $T^*$ remains of the same crossover type and does not follow resistivity behavior below $T_{SM}$ (either increase or decrease), suggesting that it is an independent feature. At doping close to optimum, $x_{opt} \approx$ 0.07, the features due to structural/magnetic transition are completely suppressed [in both $\rho_a (T)$ and $\rho _c (T)$], and the temperature dependence of the inter-plane resistivity is dominated by the maximum at $T^*$ and superconducting transition.

At the highest doping shown in Fig.~\ref{Co-rhoc-rhoa}, $x$=0.166, when the superconductivity is suppressed, a new feature appears in the temperature-dependent interplane resistivity: a shallow resistivity minimum appears at $T_{\rm CG} > T^*$.
In  Fig.~\ref{Co-rhoc-rhoa_heavy} we present the evolution of the resistivity for higher Co concentrations, starting from those on the over-doped side of the superconducting dome, $x$=0.127. The top panel shows the inter-plane resistivity, the bottom panel shows the in-plane resistivity, which shows metallic behavior for all compositions. Cross-arrows in the top  panel show the position of the high-temperature cross-over from the metallic to non-metallic temperature dependence of the inter-plane resistivity at $T_{\rm CG}$. Cross-arrows in the bottom panel show the same characteristic temperatures with respect to the temperature-dependent in-plane resistivity finding no discernible features in the $\rho _a (T)$ curves.

\begin{figure}
	
		\includegraphics[height=8.5cm,width =8.5cm]{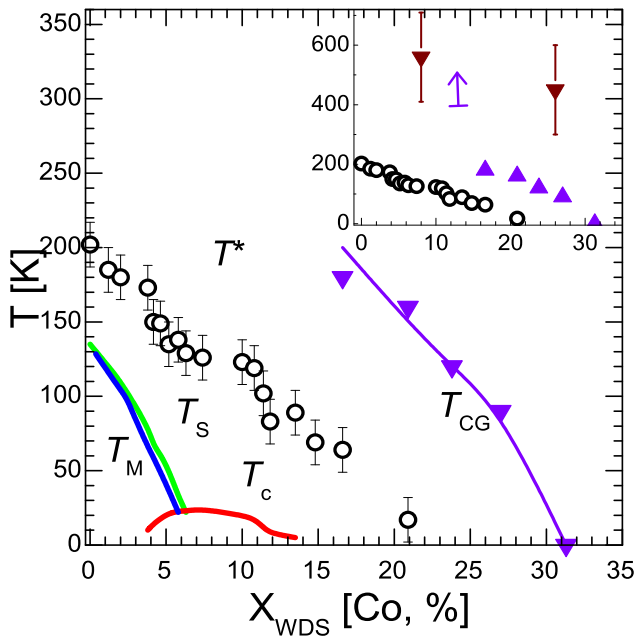}
	\caption{Temperature-doping phase diagram of Ba(Fe$_{1-x}$Co$_x$)$_2$As$_2$ as determined from inter-plane resistivity measurements. Inset shows comparison of $T^*$ and $T_{\rm CG}$, corresponding to the maxima amd minima in $\rho_ c (T)$, with $T_{PG}$ as found in NMR study by fiting the temperature-dependent Knight shift \cite{BaCoNMRPG2,BaCoNMRPG4}. Arrow in the inset shows a minimum estimate for $T_{\rm CG}$ for the border composition $x$=0.127.
}
	\label{phaseDCo}
\end{figure}

We summarize the doping evolution of the main features of the temperature-dependent resistivity in the phase diagram, Fig.~\ref{phaseDCo}. The lines of the superconducting, $T_c$, structural, $T_S$ and magnetic, $T_M$ transitions are discernible in both in-plane \cite{NiNiCo} and inter-plane resistivity. The lines corresponding to maxima, $T^*$, and minima $T_{\rm CG}$ of the temperature-dependent inter-plane resistivity find no correspondence in the temperature dependence of the in-plane transport.

This phase diagram suggests existence of a critical concentration, at which charge gap vanishes. Interestingly enough, at the concentration close to critical, $x_{\rm CG} \simeq$ 0.30, the inter-plane resistivity shows a linear temperature dependence over a broad temperature range, as seen for a sample with $x$=0.313 (red curve in the top panel of Fig.~\ref{Co-rhoc-rhoa_heavy}) for $T >$20~K. For higher $x$, the temperature-dependent resistivity develops positive curvature, and can be reasonably described by a sum of $T$-linear and $T^2$ contibutions, $\rho _c (T) = \rho_0 + AT +BT^2$, similar to in-plane transport \cite{NDL}, as shown in the top panel of Fig.~\ref{Co-rhoc-rhoa} for samples with $x$=0.343 and $x$=0.370.

\begin{figure}
	
	\includegraphics[width=1.0\linewidth]{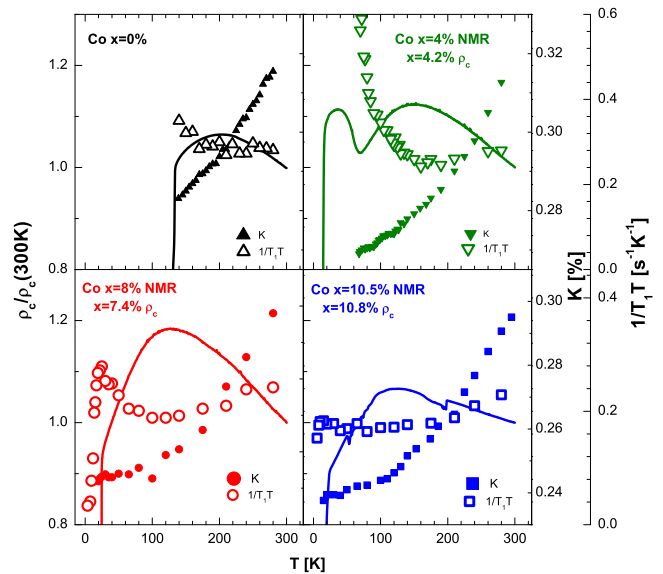}
	\caption{Comparison of the temperature-dependent interplane resistivity $\rho_c $ (solid lines, left scale) with the temperature-dependent $^{59}$Co NMR Knight shift $K(T)$ (solid symbols) and relaxation rate, $1/T_1T$, (open symbols) from Ref.~\cite{BaCoNMRPG2} (two right scales) for BaFe$_2$As$_2$ (top left panel) and Ba(Fe$_{1-x}$Co$_x$)$_2$As$_2$ with x=0.04 (top right panel), x=0.074-0.08 (bottom left panel) and x=0.105-0.108 (bottom right panel). A broad maximum in the temperature dependence of the inter-plane resistivity clearly correlates with pseudogap features in NMR measurements: a crossover slope change in $K(T)$ and the onset of a low-temperature rapid rise in $1/T_1T$. }
	\label{Co_rhoc_NMR}
\end{figure}

In Fig.~\ref{Co_rhoc_NMR} we compare the inter-plane resistivity with earlier evidence of the pseudogap in the electron-doped iron arsenides: the temperature dependence of the $^{59}$Co NMR Knight shift $K$ and $T$-normalized NMR relaxation rate, $\frac{1}{T_1T}$, as measured in Ref.~\onlinecite{BaCoNMRPG2}. We recall that, in a simple metal, both $K$ and $\frac{1}{T_1T}$ should be temperature independent. In contrast, both Knight shift and relaxation rate data in Ba(Fe$_{1-x}$Co$_x$)$_2$As$_2$ are strongly temperature-dependent. In the parent compound, $x$=0, $K(T)$ shows an increase with temperature (seen in all compositions), with a mild slope change around 210~K. On the other hand, $\frac{1}{T_1T}$ slightly increases on cooling below 200~K on approaching the temperature of the coupled structural-magnetic transition, $T_{SM}$= 135~K. These two features in $K(T)$ and $\frac{1}{T_1T}$ vs $T$ are close in temperature to a shallow maximum in $\rho_c (T)$ at around 200~K, preceding a sharp drop of resistivity at $T_{SM}$.

This correlation between the features in the temperature-dependent NMR Knight shift and the inter-plane electrical resistivity becomes clearer with increasing Co doping. The slope change in the Knight shift becomes more pronounced and, for a composition with $x$=0.105, it shifts down to $\sim$100~K. In both NMR measurements and in the inter-plane resistivity the features remain of a broad crossover type, with difficult to define characteristic temperatures. The resistivity maximum is a better defined feature, though it is still broad and its location for several samples studied for each composition could be slightly affected by the admixture of the in-plane resistivity. This admixture may affect the $\rho_c(T)$ for the $x$=0.108 sample in Fig.~\ref{Co_rhoc_NMR}, as suggested by its slight deviation from the series evolution with doping, top panel in Fig.~\ref{Co-rhoc-rhoa}. (Small jumps in the temperature dependence are extrinsic and are caused by sample cracking during thermal cycle.)

The linearly increasing NMR Knight shift \cite{BaCoNMRPG3} reflects an unusual linear temperature-dependent static magnetic susceptibility, $\chi (T)$, \cite{chen,chi1,chi2}. This anomalous linear $\chi (T)$ dependence was shown to go away  at high doping levels, being replaced by a Curie-Weiss behavior of susceptibility \cite{chen}. The magnitude of the Knight shift variation also diminishes for over-doped compositions, and it was suggested that the pseudo-gap feature disappears at critical concentration for superconductivity $x_{\rm SC} \approx$0.2 \cite{BaCoNMRPG4}.

\begin{figure}
	
		\includegraphics[width =0.93\linewidth]{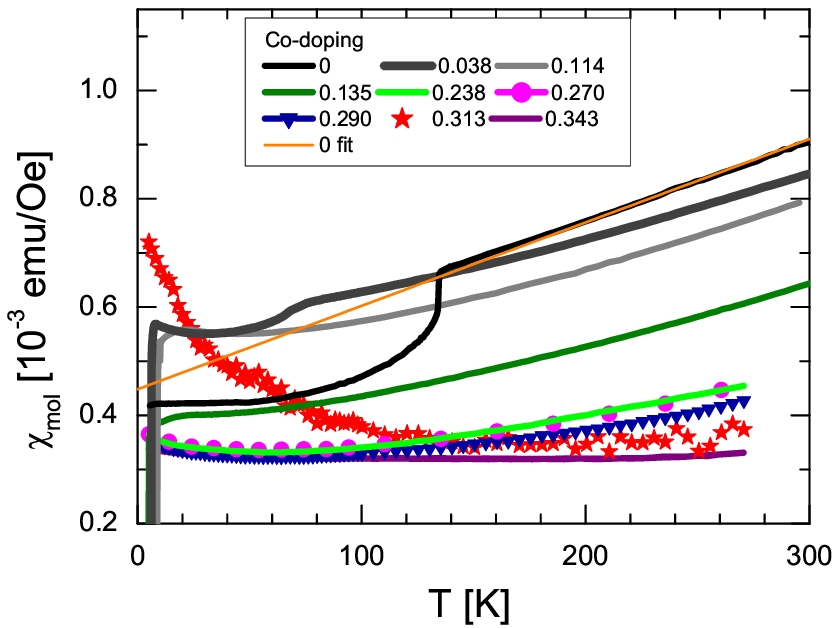}
		\includegraphics[width =0.9\linewidth]{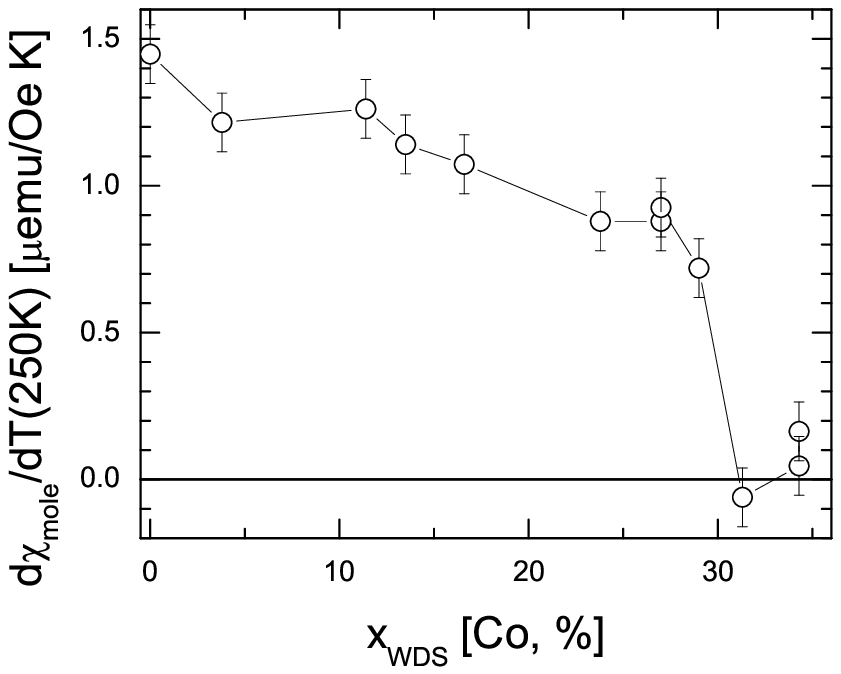}		
	\caption{(Color online) Top panel. Temperature-dependent molar magnetic susceptibility, $\chi_{mole} (T)$, measured in magnetic field $H~ \bot c$=~5~T.  The data for low dopings are from Ref.~\onlinecite{NiNiCo}. On doping increase  the slope of the $T$-linear increase in $\chi (T)$ (shown with orange line for pure composition $x$=0 at $T$=250~K) decreases and for $x$=0.313 the dependence becomes flat at $T > \sim$150~K.  Bottom panel. Doping evolution of the slope of $\chi _{mole} (T)$, $d \chi _{mole} (T) /dT$, at $T$=250~K.
}
	\label{magnetization}
\end{figure}

We performed magnetization measurements on Co doped compositions, as shown in Fig.~\ref{magnetization}. The behavior at low Co-dopings $0 <x \leq 0.166$ was studied systematically by Ni et al. \cite{NiNiCo} for both orientations of magnetic field parallel and perpendicular to the tetragonal $c$-axis and found small anisotropy. Our new measurements for $x \geq$0.166 concentrate on doping evolution of susceptibility in configuration $H \bot c$. We performed measurements with $H \parallel c$ only for sample $x$=0.343 and found no anisotropy. As can be seen from Fig.~\ref{magnetization}, the slope of the $T$-linear portion in $\chi (T)$ gradually diminishes with $x$. The
$\chi _{mole} (T)$ curve for $x$=0.290 shows very small but still clearly discernible increase with $T$, though with the slope notably smaller than the slope for $x$=0.270. For $x$=0.313 the increase of magnetic susceptibility with temperature is completely gone. Instead, $\chi (T)$ becomes temperature independent above 150~K. The Curie-Weiss increase of $\chi(T)$ on cooling at low temperatures for sample with $x$=0.313 is most likely extrinsic, it is not observed for samples with lower and higher doping, $x$=0.290 and $x$=0.343. On the other hand,  the $\chi _{mole} (T)$ dependence does not reveal any increase at high $T$ for both $x$=0.313 and two different samples of $x$=0.343.

To quantify this evolution, in the bottom panel of Fig.~\ref{magnetization} we show a slope of $\chi _{mole} (T)$ at $T$=250~K as a function of $x$. The dependence shows a dramatic change between $x$=0.290 and $x$=0.313, the same concentration as $x_{\rm CG}$ determined from inter-plane resistivity.
We note that since the crystals used in the inter-plane resistivity and magnetization measurements are from the same batches, there is minimal uncertainty in the concentration comparison. Both inter-plane resistivity and magnetic susceptibility show pronounced changes of behavior between $x$=0.290 and $x$=0.313. For $x > $0.313, the inter-plane resistivity increases monotonically and superlinearly with temperature as expected for a metal. The flat temperature-independent magnetic susceptibility is expected for Pauli susceptibility of a metal as well. The disappearence of the linearly rising $\chi (T)$ for heavily doped compositions is similar to the early report \cite{chen}, though with notable difference in the concentration boundaries ($x$=0.125 in our notations vs 0.30). This discrepancy may be a result of poor composition control in early crystals.
We would like to point out that at similar concentration $\sim$0.4 (there is no systematic doping-evolution study in the range) Hall constant becomes temperature-independent, again in line with expectation of the behavior of a  usual metal \cite{Halloverdoped}.

\section{Discussion}

\subsection{Doping evolution of the anisotropy}

The electronic structure of Co-doped BaFe$_2$As$_2$ is now well established to be three-diemnsional by various techniques \cite{3D,3D1,3D2,3D3,3D4}. However, evolution of the anisotropy with doping was never studied in a systematic way. From Fig.~\ref{rhoa300K} we can see that doping in the range from $x$=0 to $x$=0.48 leads to an approximately 3 times decrease of the in-plane resistivity at room temperature, agreeing, within error bars, with the previous measurements \cite{Alloul}. For $\rho_c$, Fig.~\ref{rhoc300K}, the variation is approximately of the same magnitude, keeping in mind an uncertainty of the factor of 2 for $\rho_c$ values. This suggests that the evolution of the anisotropy in a broad doping range is very gradual, and puts an upper bound of about a factor of two for the anisotropy change. Comparison of these numbers with the existing band structure calculations \cite{Singh,Hirschfeld,Elfimov,Athena,anisotropy,anisotropy2} should be taken with a grain of salt, since variation of the position of the As atom in the lattice from the one obtained in experiment to the one calculated from total lattice energy minimum  \cite{anisotropy,anisotropy2} brings the effect which by far exceeds the total anisotropy variation. A general decrease of both in-plane and inter-plane resistivities with doping is suggestive that charge is actually donated into the system, which does not go in line with suggestions that all doping can be treated as additional scattering \cite{Elfimov}.

The anisotropy at low temperatures, important for the anisotropy of the upper critical field in the superconducting state, is heavily affected by the structural/magnetic transition and by the pseudogap. We will discuss these effects in the next sections.

\subsection{Structural/magnetic ordering and inter-plane resistivity}

Magnetic ordering below $T_{SM}$, presumably of spin density wave (SDW) nature \cite{Mazin_SDW}, reconstructs the Fermi surface, opening a superzone gap on electron and hole pockets. This is seen as an increase of the in-plane resistivity in BaCo122 with $x$=0.025 to 0.058. The parts of the Fermi surface which are not affected by the superzone (SDW) gap, enjoy a notably reduced inelastic scattering in the magnetically ordered phase \cite{WenHall,Alloul,detwinning}. In the parent compound, in which elastic scattering is small, this decrease of scattering overcomes the loss of the carrier density so that the total conductivity increases below $T_{SM}$. Since the inter-plane transport is determined by the warped parts of the Fermi surface \cite{anisotropy}, least affected by the SDW superzone gap, the inter-plane resistivity should be affected much less by the SDW gap opening than $\rho _a$. This is indeed seen, in BaCo122, most clearly  for sample with $x$=0.012. Here, the in-plane resistivity shows an intermediate behavior between pure and heavier doped compositions: whereas $\rho _a (T)$ increases immediately below $T_{SM}$ and then shows a shallow decrease to a much higher residual value than in pure samples, the inter-plane resistivity does not manifest a local maximum below $T_{SM}$ and the resistivity decrease is almost as steep as in pure crystals. The features in the temperature-dependent resistivity upon crossing structural and magnetic transitions \cite{NiNiCo,Pratt} can be similarly resolved in in-plane and inter-plane transport, though structural transition is always less pronounsed in $\rho _c(T)$.

\subsection{Maximum in the temperature-dependent inter-plane resistivity at $T^*$}

The decrease of the inter-plane resistivity below $T^*$ shows a clear correlation with the NMR Knight shift, therefore we need to look for a common origin. An important observation is that in BaCo122 with $x$=0.08, Fig.~\ref{Co_rhoc_NMR},  both the Knight shift and the inter-plane resistivity at low temperatures, $T < T^*$, follow expectations of a metal with temperature-independent density of states: the resistivity shows metallic decrease on cooling, and the Knight shift is temperature-independent. Simultaneously, the $\frac{1}{T_1 T}$ increase indicates slowing down of magnetic fluctuations. This suggests that magnetic correlations may play an important role in the anomalies in all three measurements. Same trend hold for sample with $x$=0.105, however, the features in NMR measurements fade away with overdoping.

In the NMR study of Ref.~\onlinecite{BaCoNMRPG1}, the temperature-dependent Knight shift was fit using a two-component model, with $K=A+B*exp(-T_{PG}/T)$. At temperatures $T\ll T_{PG}$ this crosses-over to a metallic behavior with constant Knight shift detemined by the $A$ term unaffected by the pseudo-gap. At $T>T_{PG}$ both terms become temperature-independent and we can expect restoration of the metallic behavior with $K=A+B$. Fitting the temperature-dependent Knight shift, the authors determined $T_{PG}$=560~K $\pm$ 150~K for optimally doped BaCo122 samples, $x$=0.08 (with A=0.715\% and B=0.244\%), Ref.~\onlinecite{BaCoNMRPG1}, and $T_{PG}$=450~K for $x$=0.26 (with A=0.20\% and B=0.23\%). Assuming that the A and B coefficients represent partial DOS contributions of the ungapped and gapped parts of the Fermi surface, respectively, we would expect that at temperatures of the order of $T_{PG}$/3 or so, which would give us a temperature in 100 to 200~K range for optimal doping, we restore metallic resistivity temperature dependence, while resistivity decrease with temperature would be observed at higher temperatures due to carrier activation. This is consistent within general trend in evolution of $\rho _c(T)$, but not $\rho _a(T)$. This would suggest that the pseudogap affects predominantly the most warped parts of the Fermi surface.

We need to notice though, that it is difficult to explain resistivity decrease above $T^*$ solely by the existence of a {\it spin gap}, as probed by NMR \cite{BaCoNMRPG3}. Activation of spin fluctuations in the metallic phase can only increase scattering of charge carriers, which is seen in in-plane transport. Decrease of the inter-plane resistivity, despite being very small, would require rather an increase of the carrier density by excitations over a {\it charge gap}.

\subsection{ Minimum in the temperature-dependent inter-plane resistivity and charge gap}

The importance of charge gap formation for non-metallic temperature dependence of $\rho _c$ above $T^*$ is most clearly seen from the temperature-dependent inter-plane resistivity for BaCo122 composition with $0.166 \leq x \leq 0.270$. Here the resistivity shows metallic temperature dependence at high temperatures and then crosses over to a non-metallic increase below a temperature of a charge gap formation $T_{\rm CG}$. The monotonic evolution of the curves suggests that for lower dopings, $x < $0.166, the $T_{\rm CG}$ goes above the experimentally accessible range. If this is true, the end of the temperature range of monotonic resistivity decrease on heating gives us an estimate for a minimum value of charge gap for compositions with $x \leq $0.127 as above 400~K. In inset in Fig.~\ref{phaseDCo}, we compare experimentally determined $T_{\rm CG}$ from the inter-plane resistivity measurements with $T_{PG}$ determined from fitting $K(T)$ curves. Both measurements have very big error bars, and yet they do not match well.  This may suggest another possibility, that a metallic temperature dependence of $\rho_ c$ at high temperatures is only confined in some range of dopings.

The scenario with the existence of a semi-metallic charge gap was invoked for an explanation of the $T$-linear magnetic susceptibility, with simultaneous strong temperature-dependent Hall and Seebeck coefficients \cite{Sales,Alloul,mundeouk}.  In this model thermal activation of carriers over a narrow gap results in a carrier density increase with temperature. This would naturally lead to a decrease of the inter-plane resistivity with temperature. We should notice though that the magnitudes of the effects, necessary to explain temperature-dependent magnetization, by far exceed the magniture of the $\rho_c$ variation observed in our experiments. This is also true with respect to doping evolution of the characteristic temperatures, $T_{PG}$ and $T_{\rm CG}$, Fig.~\ref{phaseDCo}, which do not connect gradually.
Simultaneously the linear rise in magnetization with temperature does not coincide with resistivity maximum in our study, especially for pure BaFe$_2$As$_2$.

Despite this clear discrepancy between the two suggested explanations for the temperature-dependent magnetization and transport and our data, the effects in the inter-plane resistivity and in the magnetization are clearly correlated. In addtion, determination of the characteristic pseudogap temperature $T_{PG}$ in NMR measurements is heavily model dependent, whereas the minimum in the temperature dependence of the inter-plane resistivity, despite being boad, is rather well defined for $x >$0.166 and shows a systematic evolution. This may imply that we need to invoke different mechanisms for the explanation of the pseudogap features.

\subsection{Critical concentration}

We now turn to the evolution of $\rho _c (T)$ and magnetic susceptibility in the vicinity of $x_{\rm CG}$. The most remarkable observation here is that at $x$=0.313 the resistivity is faily linear over a broad temperature range from approximately 400~K down to 20~K and saturates at lower temperatures. For $x$=0.290 the dependence is also close to $T$-linear with a shallow slope change at $\sim$150~K. For doping with $x > 0.313$ the $\rho _c (T)$ becomes superlinear, similar to $\rho _a (T)$, and its inelastic part can be reasonably decribed as a sum of $T$-linear and $T^2$ terms, as shown for the curves $x$=0.343 and $x$=0.370. In general, evolution of $\rho _c (T)$ with doping is reminescent of the one found in systems on the verge of magnetic order and assigned to the existence of the magnetic quantum critical point. This observation suggests that the pseudogap is magnetic in origin, and is accompanied by the charge gap, rather than the charge gap itself being responsible for anomalous electronic properties.

On note, none of the anomalies in the magnetic properties is clearly reflected in the in-plane transport. This unusual single-axis effect of the pseudogap on the resistivity suggests that the magnetic action is concentrated on a small fraction of the Fermi surface, and importantly, on the most warped part contributing mainly to the inter-plane transport.
\subsection{Origin of the pseudogap}

The existence of two additional crossover lines in the phase diagram of Ba(Fe$_{1-x}$Co$_x$)$_2$As$_2$, as revealed by the inter-plane resistivity, raises an interesting question about their origins. Strong anisotropy of the pseudogap makes a scenario, in which the gap is due to superconducting pairing of charge carriers, however without superconducting condensate formation,  very unlikely. We therefore should consider the possibility that the pseudogap is arising due to either short range, or short-lived, magnetic correlations, or represents a partial gap in the electronic structure.

The magnetic structure of parent BaFe$_2$As$_2$ is characterized by a stripe-type antiferromagnetic ordering, in which antiferromagnetic spin arrangement is typical for directions both in the plane and between the planes, introducing three-dimensional magnetic Brillouin zone, poorly matching the Fermi surface. It is difficult to expect pronounced anisotropy for this case. On the other hand, if correlations seen by the inter-plane transport were the same as those of the ordered phase, it would be difficult to explain a pre-transition {\it decrease} of resistivity below $T^*$ in Co-doped samples with $x$=0.037 to 0.058, with successive {\it increase} of resistivity below $T_{SM}$. This may suggest that uniaxial anisotropy of the pseudo-gap comes from magnetic fluctuations with a different characteristic wavevector. A situation like this, when fluctuations and ordering wavevector are not the same was found in some intermetallic and heavy fermion systems \cite{different_ordering}. Indirect evidence for such a possibility comes from the fact that in a closely related EuFe$_2$As$_2$, antiferromagnetic ordering of Eu moments happens between the planes, while the Fe layer moments remain parallel in the planes \cite{Eu_magnetism,Eu_magnetism2}. Since this ordering is mediated by RKKY interactions via conduction electrons, it suggests that the generalized spin susceptibility may have maxima, which correspond to the existence of the inter-plane nesting.

In closely related BaMn$_2$As$_2$, magnetism is of local moment type, and the magnetic order is of usual AF G-type \cite{BaMn2As2-magneticstructure}. In EuRh$_2$As$_2$ , commensurate and incommensurate spiral-like structures with propagation along the $c$-axis are found. \cite{EuRh2As2} Although these compounds are differing in band structure and electron count, these observations of different types of ordering may be suggesting that various magnetic structures are not very different in energy.

In the lack of any evidence for the existence of such correlations in Co-doped BaFe$_2$As$_2$, we just speculate what consequences uniaxial character of the pseudogap may have. This type of a pseudogap is impossible in two dimensional cuprates, it is a direct consequence of the difference in the dimensionality of the electronic and magnetic systems in the cuprates and in iron arsenides. If the link between the symmetry of the pseudogap and of the superconducting order parameter, as found in the cuprates \cite{Ronning,Kaminski},  is preserved in the iron arsenides, $c$-axis pseudogap would correspond to a superconducting gap having maxima/minima at the poles. This scenario was invoked theoretically for explanation of unusual behavior in the superconducting gap \cite{Laad-Graco}. In experiment, variation of the superconducting gap with polar angle is found in inelastic neutron scattering in Ni doped compound at optimal doping \cite{resonanceNi}, with gap magnitude decreasing towards the poles, and in penetration depth study of BaNi122 \cite{CatalinNi}. It is directly revealed in the inter-plane heat transport study \cite{c-axis-thermalcond}, as opposed to the in-plane study \cite{Coheatinplane}.

Finally we would like to point to a certain similarity in the critical behaviour of the inter-plane resistivity in BaCo122 and in CeCoIn$_5$. In CeCoIn$_5$, a true critical behavior at a field-tuned QCP \cite{Paglione,Bianchi} with $T$-linear resistivity and violation of the Wiedemann-Franz law is observed for transport along the tetragonal $c$-axis \cite{Science}. Transport in the plane perpendicular to $c$-axis, despite showing unusual power law behavior, obeys the WF law \cite{nonvanishing}.

\section{Conclusion}

Contrary to the in-plane electrical resistivity, which away from the domain of structural/magnetic ordering shows monotonic metallic temperature dependence, inter-plane resistivity, $\rho _c(T)$,  reveals anomalous features clearly correlating with features in the temperature dependence of the the NMR Knight shift and spin-relaxation rate, assigned to the formation of the pseudo-gap. Evolution of $\rho _c(T)$ with doping reveals two characteristic energy scales, of the the resistivity maximum (seen for compositions $0 \leq x < \sim 0.2$) and resistivity minimum at a tempearture $T_{\rm CG}$, seen for $0.166 \leq x <  x_c$, $x_c \approx $0.3. The temperature-dependent $\rho _c$ is close to linear close to $x_{\rm CG}$ and super-linear for $x>x_{\rm CG}$. None of these features are evident in the in-plane resistivity $\rho_a(T)$. For doping levels $x <x_{\rm CG}$, $\chi (T)$ shows a known, anomalous, $T$-linear dependence, which disappears for $x>x_{\rm CG}$. These features are consistent with the existence of a uniaxial charge gap, accompanying formation of the magnetic pseudogap, and its critical suppression with doping. This evolution suggests existence of critical point for pseudogap order. The superconducting dome is confined inside the pseudo-gap dome.


\section{Acknowledgements}

We thank A. Kaminski and A. Kreyssig for useful discussions. Work at the Ames Laboratory was supported by the Department of Energy-Basic Energy Sciences under Contract No. DE-AC02-07CH11358. R. P. acknowledges support from Alfred P. Sloan Foundation.



\begin{references}

\bibitem{BandM} J. G. Bednorz and K. A. Muller, J. G. Bednorz and K. A. Muller, Rev. Mod. Phys. {\bf 60}, 585
(1988).
\bibitem{TS} T. Timusk, and B. Statt, Rep. Prog. Phys. {\bf 62}, 61 (1999).
\bibitem{Kaminski} T.~Kondo, R.~Khasanov, T.~Takeuchi, J.~Schmalian, and A.~Kaminski , Nature {\bf 457}, 296 (2009).

\bibitem{pseudogap_origin} M.~R.~Norman, D.~Pines, and C.~Kallin,
Adv. Phys. {\bf 54}, 715 (2005).


\bibitem{magnetism} V. Hinkov, P.~Bourges, S.~Pailhes, Y.~Sidis, A.~Ivanov, C.~D.~Frost, T.~G.~Perring, C.~T.~Lin, D.~P.~Chen, and B.~Keimer, Nature Phys. {\bf 3}, 780 (2007).

\bibitem{pairing}
 K.~K.~Gomes, A.~N.~Pasupathy, A.~Pushp, S.~Ono, Y.~Ando, and A.~Yazdani, Nature {\bf 447}, 569 (2007).


\bibitem{greene} N.P. Armitage, P. Fournier, and R.L. Greene, arXiv:0906.2931

\bibitem{hosono} Y. Kamihara, T. Watanabe, M. Hirano, and H. Hosono, J. Am. Chem. Soc. {\bf 130}, 3296 (2008).

\bibitem{Mazin-Nature} I.~I.~Mazin, Nature {\bf 464}, 183 (2010).

\bibitem{PG1111_Hall_resistivity} Y. Kohama, Y. Kamihara, S. A. Baily, L. Civale, S. C. Riggs, F. F. Balakirev, T. Atake, M. Jaime, M. Hirano, and H. Hosono,
 Phys. Rev. B {\bf 79}, 144527 (2009).


\bibitem{PG1111_NMR}
K. Ahilan, F. L. Ning, T. Imai, A. S. Sefat, R. Jin, M. A. McGuire,
B. C. Sales, and D. Mandrus, Phys. Rev. B {\bf 78}, 100501(R) (2008).

Y.~Nakai, S.~Kitagawa, K.~Ishida, Y.~Kamihara, M.~Hirano, and H.~Hosono,   New J. Phys. {\bf 11}, 045004 (2009).

\bibitem{PG1111_Photoemission1} Y. Ishida, T. Shimojima, K. Ishizaka, T. Kiss, M. Okawa, T. Togashi, S. Watanabe, X.-Y. Wang, C.-T. Chen, Y. Kamihara, M. Hirano, H. Hosono, and S. Shin,
Phys. Rev. B {\bf 79}, 060503 (2009).

\bibitem{PG1111_Photoemission2}
T.~Sato, S.~Souma, K.~Nakayama, K.~Terashima, K.~Sugawara, T.~Takahashi, Y.~Kamihara, M.~Hirano, and H.~Hosono,
J. Phys. Soc. Jpn. {\bf 77}, 063708 (2008).

\bibitem{Sm_Russian} A.L. Solovjov, S.L. Sidorov, Yu.V. Tarenkov, and A.I. D'yachenko, arXiv:1002.2306

\bibitem{ARPES_PGBaK}
Y.-M. Xu, P. Richard, K. Nakayama, T. Kawahara, Y. Sekiba, T. Qian, M. Neupane, S. Souma, T. Sato, T. Takahashi, H. Luo, H.-H. Wen, G.-F. Chen, N.-L. Wang, Z. Wang, Z. Fang, X. Dai, and H. Ding,
arXiv: 0905.4467.

\bibitem{ishida} K. Ishida, Y. Nakai, H. Hosono,
J. Phys. Soc. Jpn. {\bf 78}, 062001 (2009).


\bibitem{BaCoNMRPG1}
F.L. Ning, K. Ahilan, T. Imai, A. S. Sefat, R. Jin, M. A. McGuire, B. C. Sales, and D. Mandrus
J. Phys. Soc. Jpn. {\bf 77}, 103705 (2008).

\bibitem{ARPESBaCo} Y. Sekiba, T. Sato, K. Nakayama, K. Terashima, P. Richard, J. H. Bowen, H. Ding, Y.-M. Xu, L. J. Li, G. H. Cao, Z.-A. Xu, and T. Takahashi,
New J. Phys. {\bf 11},  025020 (2009).

\bibitem{NiNiCo} N. Ni, M. E. Tillman, J.-Q. Yan, A. Kracher, S. T. Hannahs, S. L. Bud'ko, and P. C. Canfield, Phys. Rev. B {\bf 78}, 214515 (2008).

\bibitem{NDL} N.~Doiron-Leyraud, P.~Auban-Senzier, S. René de Cotret, C.~Bourbonnais, D.~Jérome, K.~Bechgaard, and L.~Taillefer, Phys. Rev. B {\bf 80}, 214531 (2009).


\bibitem{anisotropy} M. A. Tanatar, N. Ni, C. Martin, R. T. Gordon, H. Kim, V. G. Kogan, G. D. Samolyuk, S. L. Bud'ko, P. C. Canfield, and R. Prozorov, Phys. Rev. B {\bf 79}, 094507 (2009).
\bibitem{anisotropy2} M. A. Tanatar, N. Ni, G. D. Samolyuk, S. L. Bud'ko, P. C. Canfield, and R. Prozorov, Phys. Rev. B {\bf 79}, 134528 (2009).

\bibitem{detwinning} M. A. Tanatar, E. C. Blomberg, A. Kreyssig, M. G. Kim, N. Ni, A. Thaler,
S. L. Bud'ko, P. C. Canfield, A. I. Goldman, I. I. Mazin, and R. Prozorov, Phys. Rev. B {\bf 81}, 184508 (2010).


\bibitem{BaCoNMRPG2}
F.L. Ning, K. Ahilan, T. Imai, A. S. Sefat, R. Jin, M. A. McGuire, B. C. Sales, and D. Mandrus,
J. Phys. Soc. Jpn. {\bf 78}, 013711 (2009).

\bibitem{Mathur} N. D. Mathur, F. M. Grosche, S. R. Julian,  I. R. Walker, D. M. Freye, R. K. W. Haselwimmer, and G. G. Lonzarich,
Nature {\bf 394}, 39 (1998).




\bibitem{SUST} M. A. Tanatar, N. Ni, S. L. Bud'ko, P. C. Canfield, and R. Prozorov, Supercond. Sci. Technol. {\bf 23}, 054002 (2010).


\bibitem{vortex} R. Prozorov, N. Ni, M. A. Tanatar, V. G. Kogan, R. T. Gordon, C. Martin, E. C. Blomberg, P. Prommapan, J. Q. Yan, S. L. Bud'ko, and P. C. Canfield, Phys. Rev. B {\bf 78}, 224506 (2008).

\bibitem{c-axis-thermalcond} J.-Ph. Reid, M. A. Tanatar, H. Shakeripour, X. G. Luo, N. Doiron-Leyraud,
N. Ni, S. L. Bud'ko, P. C. Canfield, R. Prozorov, and L.~Taillefer,
arXiv:1004.3804.


\bibitem{BETSGaCl4}
 M. A. Tanatar, T. Ishiguro, H. Tanaka, and H. Kobayashi,
Phys. Rev. B {\bf 66}, 134503 (2002).

\bibitem{Alloul} F. Rullier-Albenque, D. Colson, A. Forget, and H. Alloul
Phys. Rev. Lett. {\bf 103},057001 (2009).

\bibitem{mundeouk}
E.~D.~Mun, S.~L.~Bud'ko, N.~Ni, A.~N.~Thaler, and P.~C.~Canfield
Phys. Rev. B {\bf 80}, 054517 (2009).



\bibitem{Kaminskiivnature}
C.~Liu, T.~Kondo, R.~M.~Fernandes, A.~D.~Palczewski, E.~D.~Mun, N.~Ni, A.~N.~Thaler, A.~Bostwick, E.~Rotenberg, J.~Schmalian, S.~L.~Bud'ko, P.~C.~Canfield, and A.~Kaminski, Nature Phys. (accepted), arXiv:0910.1799




\bibitem{BaCoNMRPG4} F. L. Ning, K. Ahilan, T. Imai, A. S. Sefat, M. A. McGuire, B. C. Sales, D. Mandrus, P. Cheng, B. Shen, and H.-H Wen, Phys. Rev. Lett. {\bf 104}, 037001 (2010).

\bibitem{BaCoNMRPG3} K.~Ahilan, F.~L.~Ning, T.~Imai, A.~S.~Sefat, M.~A.~McGuire, B.~C.~Sales, D.~Mandrus, P.~Cheng, B.~Shen, and H.~H.~Wen, arXiv:0910.1071

\bibitem{chen}  X. F. Wang, T. Wu, G. Wu, H. Chen, Y. L. Xie, J. J. Ying, Y. J. Yan, R. H. Liu, and X. H. Chen, New J. Phys. {\bf 11}, 045003  (2009).

\bibitem{chi1} J.-Q. Yan, A. Kreyssig, S. Nandi, N. Ni, S. L. Bud'ko, A. Kracher, R. J. McQueeney, R. W. McCallum, T. A. Lograsso, A. I. Goldman, and P. C. Canfield,
 Phys. Rev. B {\bf 78}, 024516 (2008).

\bibitem{chi2}R. Klingeler, N. Leps, I. Hellmann, A. Popa, U. Stockert, C. Hess, V. Kataev, H.-J. Grafe, F. Hammerath, G. Lang, G. Behr, L. Harnagea, S. Singh, B. Buechner,
Phys. Rev. B {\bf 81}, 024506 (2010).



\bibitem{Halloverdoped} N.~Katayama, Y.~Kiuchi, Y. Matsushita, and K.~Ohguchi, J. Phys. Soc. Jpn. {\bf 78}, 123702 (2009).




\bibitem{3D} W. Malaeb, T. Yoshida, A. Fujimori, M. Kubota, K. Ono, K. Kihou, P. M. Shirage, H. Kito, A. Iyo, H. Eisaki, Y. Nakajima, T. Tamegai, and R. Arita,
J. Phys. Soc. Jpn. {\bf 78}, 123706 (2009).

\bibitem{3D1} P. Vilmercati, A. Fedorov, I. Vobornik, U. Manju, G. Panaccione, A. Goldoni, A. S. Sefat, M. A. McGuire, B. C. Sales, R. Jin, D. Mandrus, D. J. Singh, and N. Mannella,
Phys. Rev. B {\bf 79}, 220503 (2009).

\bibitem{3D2} Y.~Sekiba, T.~Sato, K.~Nakayama, K.~Terashima, P.~Richard, J.~H.~Bowen, H.~Ding, Y.-M.~Xu, L.~J.~Li, and G.~H.~Cao, New J. Phys. {\bf 11}, 025020 (2009).


\bibitem{3D3} C.~Utfeld, J.~Laverock, T.~D.~Haynes, S.~B.~Dugdale, J.~A.~Duffy, M.~W.~Butchers, J.~W.~Taylor, S.~R.~Giblin, J.~G.~Analytis, J.-H.~Chu, I.~R.~Fisher, M.~Itou, and Y.~Sakurai,
Phys. Rev. B {\bf 81}, 064509 (2010).

\bibitem{3D4}
S.~Thirupathaiah, S.~de Jong, R.~Ovsyannikov, H.~A.~Dürr, A.~Varykhalov, R.~Follath, Y.~Huang, R.~Huisman, M.~S.~Golden, Yu-Zhong Zhang, H.~O.~Jeschke, R.~Valentí, A.~Erb, A.~Gloskovskii, and J. Fink1,
Phys. Rev. B {\bf 81}, 104512 (2010).


\bibitem{Singh} D. J. Singh, Phys. Rev. B {\bf 78}, 094511 (2008).

\bibitem{Athena}
A. S. Sefat, R. Jin, M. A. McGuire, B. C. Sales, D. J. Singh, and D. Mandrus  Phys. Rev. Lett. {\bf 101}, 117004 (2008).

\bibitem{Hirschfeld} A. F.~Kemper, C.~Cao, P.~J.~Hirschfeld, and H.-P.~Cheng
Phys. Rev. B {\bf 80}, 104511 (2009).


\bibitem{Elfimov} H. Wadati, I. Elfimov, and G. A. Sawatzky, arXiv:1003.2663


\bibitem{Mazin_SDW}
I.~I.~Mazin and M.~D.~Johannes, Nature Phys. {\bf 5}, 141 (2009).
M. D. Johannes and I. I. Mazin, Phys. Rev. B {\bf 79}, 220510 (2009).


\bibitem{WenHall} Lei Fang, Huiqian Luo, Peng Cheng, Zhaosheng Wang,
Ying Jia, Gang Mu, Bing Shen, I. I. Mazin, Lei Shan,
Cong Ren and Hai-Hu Wen, Phys.Rev.B {\bf 80}, 140508 (R) (2009).


\bibitem{Pratt} D. K. Pratt, W. Tian, A. Kreyssig, J. L. Zarestky, S. Nandi, N. Ni, S. L. Bud'ko, P. C. Canfield, A. I. Goldman, and R. J. McQueeney,
Phys. Rev. Lett. {\bf 103}, 087001 (2009).




\bibitem{Sales} B. C. Sales, M. A. McGuire, A. S. Sefat, and D. Mandrus,
Physica C {\bf 470}, 304 (2010).



\bibitem{different_ordering} See, for example, H.~Lin, L.~Rebelsky, M.~F.Collins, J.~D.~Garrett, and W.~J.~L.~Buyers,
Phys. Rev. B {\bf 43}, 13232 (1991).


\bibitem{Eu_magnetism} Y. Xiao, Y. Su, M. Meven, R. Mittal, C. M. N. Kumar, T. Chatterji, S. Price, J. Persson, N. Kumar, S. K. Dhar, A. Thamizhavel, and Th. Brueckel,
Phys. Rev. B {\bf 80}, 174424 (2009).

\bibitem{Eu_magnetism2}
J. Herrero-Martín, V. Scagnoli, C. Mazzoli, Y.~Su, R. Mittal, Y. Xiao, Th. Brueckel, N.~Kumar, S.~K.~Dhar, A. Thamizhavel, and L.~Paolasini, Phys. Rev. B {\bf 80}, 134411 (2009).

\bibitem{BaMn2As2-magneticstructure} Yogesh Singh, M. A. Green, Q. Huang, A. Kreyssig, R. J. McQueeney, D. C. Johnston, and A. I. Goldman, Phys. Rev. B {\bf 80}, 100403 (R) (2009).

\bibitem{EuRh2As2}
S. Nandi, A. Kreyssig, Y. Lee, Yogesh Singh, J. W. Kim, D. C. Johnston, B. N. Harmon, and A. I. Goldman,
Phys. Rev. B {\bf 79}, 100407 (R) (2009).

\bibitem{Ronning} F.~Ronning, C.~Kim, D.~L.~Feng, D.~S.Marshall, A.~G.~Loeser, L.~L.~Miller, J.~N.~Eckstein, I.~Bozovic, and Z.~X.~Shen, Science {\bf 282}, 2067 (1998).

\bibitem{Laad-Graco} M. S. Laad and L. Craco,
Phys. Rev. Lett. {\bf 103}, 017002 (2009).

\bibitem{resonanceNi} Songxue Ch, A. Schneidewind, Jun Zhao, L. W. Harriger, L. Li, Y. Luo, G. Cao, Zh. Xu, M. Loewenhaupt, J. Hu, and P. Dai,
Phys. Rev. Lett.{\bf 102}, 107006 (2009).

\bibitem{CatalinNi} C. Martin, H. Kim, R. T. Gordon, N. Ni, V. G. Kogan, S. L. Bud'ko, P. C. Canfield, M. A. Tanatar, and R. Prozorov, Phys.Rev.B {\bf 81}, 060505 (R) (2010).

\bibitem{Coheatinplane} M. A. Tanatar, J.-Ph. Reid, H. Shakeripour, X. G. Luo, N. Doiron-Leyraud, N. Ni, S. L. Bud'ko, P. C. Canfield, R. Prozorov, and Louis Taillefer,
Phys. Rev. Lett. {\bf 104}, 067002 (2010).


\bibitem{Paglione} Johnpierre Paglione, M. A. Tanatar, D. G. Hawthorn, Etienne Boaknin, R. W. Hill, F. Ronning, M. Sutherland, Louis Taillefer, C. Petrovic, and P. C. Canfield,
Phys. Rev. Lett. {\bf 91}, 246405 (2003)

\bibitem{Bianchi} A. Bianchi, R. Movshovich, I. Vekhter, P. G. Pagliuso, and J. L. Sarrao,
Phys. Rev. Lett. {\bf 91}, 257001 (2003).


\bibitem{Science} M.~A.~Tanatar, J.~Paglione, C.~Petrovic, and L.~Taillefer, Science {\bf 316}, 1320 (2007).


\bibitem{nonvanishing} Johnpierre Paglione, M. A. Tanatar, D. G. Hawthorn, F. Ronning, R. W. Hill, M. Sutherland, Louis Taillefer, and C. Petrovic,
Phys. Rev. Lett. {\bf 97}, 106606 (2006).













\end{references}
\end{document}